\documentclass[a4paper,11pt]{article}
\usepackage{jheppub} 
\usepackage{graphicx, graphics, color}
\usepackage[T1]{fontenc} 
\usepackage{amsmath, amssymb, bm, graphicx, graphics, color, mathrsfs}
\usepackage{epsfig}
\usepackage{dcolumn}
\usepackage{bm}
\usepackage{tikz}
\usepackage{graphicx, graphics, color}
\usepackage{dcolumn}
\usepackage{bm}
\usepackage{hyperref}
\usepackage[mathlines]{lineno}

\newcommand{\be}{\begin{equation}}
\newcommand{\ee}{\end{equation}}
\newcommand{\ben}{\begin{eqnarray}}
\newcommand{\een}{\end{eqnarray}}

\newcommand{\cO}{{\cal O}}

\newcommand{\p}{\partial}

\newcommand{\tmu}{{\tilde \mu}}

\newcommand{\ep}{\epsilon}

\newcommand{\ga}{\gamma}

\newcommand{\tB}{{\tilde B}}
\newcommand{\tE}{{\tilde E}}

\newcommand{\tga}{{\tilde \gamma}}

\newcommand{\zz}{\mathbb{Z}_2}

\title{\boldmath 
Hydrodynamics of topological Dirac semi-metals with chiral and $\mathbb{Z}_2$ anomalies}

\author[1]{Marek Rogatko\note{rogat@kft.umcs.lublin.pl, marek.rogatko@poczta.umcs.lublin.pl}}
\author[2]{Karol I. Wysokinski\note{karol@tytan.umcs.lublin.pl}}
\affiliation{Institute of Physics \\
Maria Curie-Sk{\l}odowska University \\
20-031 Lublin, pl. Marii Curie-Sk{\l}odowskiej 1, Poland}

\abstract{We consider the hydrodynamical model of topological Dirac semi-metal possessing two Dirac 
nodes separated in momentum space along a rotation axis. It has been argued that the system 
in question, except the chiral anomaly, is endowed with the other one $\zz$. In order to model such a system
we introduce two $U(1)$-gauge fields. The presence of the additional $\zz$ anomaly leads to the 
non-trivial modifications of hydrodynamical equations and to the appearance of new kinetic coefficients  
bounded with the vorticity and the magnetic parts of Maxwell and auxiliary $U(1)$-gauge fields.}


\begin{document} 

\maketitle
\flushbottom



\section{Introduction}
\label{sec:intro}
The possibility of the hydrodynamic approach to transport  
relies on the fact that strong interactions of the constituent particles which,
at low energies and long length - scales,  move  like a fluid can be described with only 
a few collective or slowly varying variables.
These include the local velocity $v(x)$, temperature $T(x)$ and chemical potentials $\mu_a(x)$ 
related to all  conserved charges (their densities are denoted by $\rho_a(x)$). The
hydrodynamics of the relativistic fluid has been developed by Landau  \cite{landau1959} and others \cite{eckart1940}
and generalized to take relativistic triangle anomalies \cite{adler1969,bell1969} 
into account.  

A purely hydrodynamic derivation of the anomaly effects, considering the first order in derivation expansion 
was presented in \cite{son09}. The idea
was to examine the local entropy production rate in the presence of anomalies and impose the positivity constraint  stemming from 
the second law of thermodynamics. It was shown that 
the contributions from the anomaly to the entropy production were locally unbounded and might
potentially violate the second law of thermodynamics, so the proper generalizations were necessary. In turn, these facts lead to 
a set of differential equations for the novel transport coefficients connected with the anomaly. Further, 
this idea was implemented to the case of anomalous superfluids \cite{lub10}-\cite{bha11}
and non-abelian symmetry \cite{eli10,nei11}. 

On the other hand, the chiral magnetic anomaly, i.e., 
anomaly induced phenomenon of electric charge separation along the axis
of the applied magnetic field in the presence of fluctuating topological charge was widely studied \cite{kha06}-\cite{kha14}.
The chiral magnetic effect in hydrodynamical approach, in a model with two conserved currents, vector 
and axial and the associated chemical potentials
was studied in \cite{isa11}.
The aforementioned phenomenon has attracted a lot of attention due to the possible explanation of an experimentally 
observed charge asymmetry in heavy ion collisions and provided explanation
for the observed decay of neutral pion into photons. 

The anomalies have been predicted \cite{nielsen1983} 
and later found \cite{son2013,burkov2015} to play an important part in the description of electrons in solids.
Recently, the hydrodynamics with vector and axial currents and the holographic charged fluid with electric separation
effect was paid attention to \cite{bu18}. It turns out that the gravitational backreaction effect may be responsible 
for the emergence of an axial current as a response to an axial electric one.


The necessity of relativistic description of electrons in solids may appear
superficial, as the velocity of electrons in solids typically equals a small fraction
of the light velocity. However, the spectrum of electrons in many materials and
close to some special points in the Brillouin zone, has a relativistic form characteristic for massless
particles. Such Dirac - like massless nature of spectrum is protected by symmetries and 
has been spotted in the two - dimensional graphene \cite{castroneto2009} and at the surfaces  
of the crystalline topological insulators \cite{hasan2010}. The Dirac - like spectrum is predicted 
 and observed in the  three - dimensional materials known 
as Dirac or Weyl semi-metals \cite{liu2014,su2015,liu2014a,lu2017,lundgren2014,neupane2014,neupane2016}. 
The transport properties of graphene 
 with the Dirac point at the Fermi energy have been proposed to follow 
the hydrodynamic description \cite{foster2009}. Later measurements confirmed the hydrodynamic
behavior of electrons in graphene \cite{crossno2016} and   in three - dimensional systems
\cite{neupane2014,neupane2016,liang2015,liang2017,he2016}. All this makes the relativistic hydrodynamic approach
to electrons in condensed matter  a timely and important issue.

Moreover, the recent experimental works provide clear evidences that chiral anomaly is observed in condensed matter systems.
Namely, it was spotted in Dirac semi-metal Na$_3$Bi \cite{xio15}, ZrTe$_5$ \cite{li16}, as well as, 
in Weyl semi-metal TaAs and NbP \cite{zha15}-\cite{goo17}.  
The mentioned two classes of Dirac semi-metals (DSM)
have acquired attention in the contemporary investigations. In the first one  
the Dirac points appear at the time reversal invariant
momenta in the first Brillouin zone, while in the other  
the Dirac points take place in pairs and are separated in momentum 
space along a rotational axis \cite{young2012,chiu2016}.
It turns out that the experimentally found examples of DSM belong 
mainly to the second class of the aforementioned materials.

The Dirac points in the second class of aforementioned semi-metals are endowed 
with a non-trivial $\mathbb{Z}_2$ topological invariant 
protecting the nodes and leading to the presence of Fermi arc surface states \cite{yan14}-\cite{kob15}. 
The novel  charge,   in a close analogy to the chiral one, is also not conserved under the action of external
fields. The non-conservation of the novel anomalous charge has been argued to have an effect 
on transport characteristics of materials \cite{burkov2016}. Thus the recent studies of three dimensional
condensed matter systems open the doors to  symmetries not spotted in other relativistic objects, making
the subject even more intriguing. 

The main motivation  behind our considerations is a natural question about the possible influence 
of $\zz$ topological charge on transport characteristics of the studied materials.
The aim of the present work is to generalize relativistic hydrodynamics
including the chiral anomaly \cite{adler1969,bell1969} and the additional anomaly,
which we call $\mathbb{Z}_2$ anomaly after the paper \cite{burkov2016}. The two anomalous charges 
in the considered theory require the existence of 
the two conjugate to them chemical potentials ($\mu$ and $\mu_d$). At the equilibrium 
both chemical potentials take zero values. Accordingly we also introduce 
 two $U(1)$-gauge fields, one being the standard Maxwell field coupled to the chiral anomalous charge
and other coupled to the $\mathbb{Z}_2$ topological charge. The derived  set of hydrodynamic equations generalizes
those previously found \cite{son09} and extensively discussed \cite{neiman2011,lucas2016} in the literature. 

The organization of the paper is as follows.
In the next section \ref{sec:model} we present the calculations leading to generalization of  
the relativistic hydrodynamic equations \cite{son09} in such a way that they take into account 
two anomalous charges, responsible for chiral and $\zz$ anomalies.
In section 3 we conclude with the discussion of the main results and possible modifications of
the transport characteristics of materials.

\section{Hydrodynamical model}
\label{sec:model}
In this section we examine the hydrodynamical model of topological Dirac semi-metal in which two Dirac nodes, 
protected by rotational symmetry, are
separated in momentum space along a rotation axis. It has been argued that the aforementioned system 
constitutes a source of the additional $\zz$ anomaly, except the chiral one,
which leads to the non-conservation of the corresponding anomalous $\zz$ topological charge \cite{burkov2016}. In order to model such a system we consider
anomalous charges connected with two $U(1)$-gauge fields. One of them is the ordinary Maxwell 
gauge field, the other is the additional one connected with the $\zz$ anomalous charge.
The hydrodynamical equations of motion 
in the presence of $\mathbb{Z}_2$ and chiral anomalies are provided by
\ben
\p_\alpha T^{\alpha \beta}(F,B) &=& F^{\beta \alpha} j_{\alpha}(F) + B^{\beta \alpha} j_{\alpha}(B),\\ \label{jf}
\p_\alpha j^\alpha (F) &=& C_1~E_{(F)\alpha} B^{(F) \alpha} + C_2~\tE_{(B)\alpha} \tB^{(B)\alpha},\\ \label{jb}
\p_\alpha j^\alpha (B) &=& C_3~\tE_{(B)\alpha} B^{(F)\alpha} + C_4~E_{(F) \alpha} \tB^{(B)\alpha},
\een
where $C_i,~i=1,\dots,4$ denote the constants which determine the adequate anomalies.
The electric and magnetic components of the two gauge fields, in the fluid rest frame, are written respectively as
\ben
E^{(F)}_{\alpha} &=& F_{\alpha \beta} u^\beta, \qquad B^{(F)}_{\alpha} = \frac{1}{2} \ep_{\alpha \beta \rho \delta}~u^\beta~F^{\rho \delta}, \\
\tE^{(B)}_{\alpha} &=& B_{\alpha \beta} u^\beta    \qquad  \tB^{(B)}_{\alpha} = \frac{1}{2} \ep_{\alpha \beta \rho \delta}~u^\beta~B^{\rho \delta}.
\een
 $F_{\alpha \beta} = 2 \p_{[ \alpha} A_{\beta ]}$ stands for the ordinary Maxwell field strength tensor, while
the second $U(1)$-gauge field $B_{\alpha \beta}$ is given by $B_{\alpha \beta} = 2 \p_{[ \alpha} B_{\beta ]}$.
On the other hand, $j_\alpha(F),~j_\alpha (B)$ represent the adequate currents connected with the gauge fields.
The relation (\ref{jf}) describes the modifications of the anomalous chiral charge conservation law
when the external magnetic and electric fields parallel to each other are applied to the system, 
while the equation (\ref{jb}) expresses the changes of the anomalous $\mathbb{Z}_2$ charge conservation law.

The energy momentum tensor and the currents needed for the hydrodynamic 
description of the relativistic fluid are given by \cite{landau1959,son09}
\ben \label{e21}
T^{\alpha \beta} &=& \Big( \ep +p \Big) u^\alpha u^\beta +  p g^{\alpha \beta} + \tau^{\alpha \beta},\\ \label{e22}
j^\alpha (F) &=& \rho~u^\alpha + V_F^\alpha,\\ \label{e23}
j^\alpha (B) &=& \rho_d~u^\alpha + V_B^\alpha,
\een
where $\ep$ is the energy per unit volume, $p$ the pressure of the fluid, $\rho$,~$\rho_d$ are the $U(1)$ 
charge densities, while $\tau^{\alpha \beta}$ 
and $V_{F(B)}^\alpha$ depict higher order corrections in velocity gradients  and  correspond to the dissipative effects
in the fluid. In the rest frame of the fluid element, there are no dissipative forces  and
$u_\alpha~\tau^{\alpha \beta }= 0$ and $u_\alpha~V_F^\alpha =u_\alpha~V_B^\alpha= 0$.
The four-vector $u^\alpha$, with the normalization $u_\alpha u^\alpha =-1$, describes the flow of the considered fluid.
 
Using the thermodynamical relations 
\be
\ep + p = T s + \mu ~\rho + \mu_d~\rho_d, \qquad d p = s~dT+ \rho~ d\mu   +  \rho_d~ d\mu_d,
\label{gibbs}
\ee
where $s$ is the entropy per unit volume, the explicit expression for energy-momentum tensor and $u_\beta~\p_\alpha T^{\alpha \beta}$, as well as, 
the expressions for $\p_\alpha j^\alpha (F)$ and $\p_\alpha j^\alpha(B)$, one arrives at the following relation:
\ben \label{ent}
\p_\alpha s^\alpha &=& \p_\alpha \Big[ s u^\alpha - \frac{\mu}{T} V_F^\alpha - \frac{\mu_d}{T} V_B^\alpha \Big] =
- \frac{1}{T} \tau^{\alpha \beta} \p_\alpha u_\beta~ \\ \nonumber
&-& V_F^\alpha~\bigg[ \p_\alpha \Big( \frac{\mu}{T} \Big) - \frac{E^{(F)}_{\alpha}}{T} \bigg]
- V_B^\alpha~\bigg[ \p_\alpha \Big( \frac{\mu_d}{T} \Big) - \frac{\tE^{(B)}_{\alpha}}{T} \bigg] \\ \nonumber
&-& \frac{\mu}{T} ~\Big( C_1~E^{(F)}_{\alpha} B^{(F)\alpha} + C_2~\tE^{(B)}_{\alpha} \tB^{(B)\alpha} \Big) 
- \frac{\mu_d}{T} ~\Big( C_3~\tE^{(B)}_{\alpha} B^{(F)\alpha} + C_4~E^{(F)}_{\alpha} \tB^{(B)\alpha} \Big),
\een
where in our system we define
\ben
V_F^\alpha &=& - \sigma_F~\Big[ T~P^{\alpha \beta}~\p_\beta \Big( \frac{\mu}{T} \Big) - E^{(F)\alpha} \Big] 
- \sigma_{F \tB}~\Big[ T~P^{\alpha \beta}~\p_\beta \Big( \frac{\mu_d}{T} \Big) - \tE^{(B)\alpha} \Big],\\
V_B^\alpha &=& - \sigma_B~\Big[ T~P^{\alpha \beta}~\p_\beta \Big( \frac{\mu_d}{T} \Big)  -\tE^{(B)\alpha} \Big] 
- \sigma_{B F}~\Big[ T~P^{\alpha \beta}~\p_\beta \Big( \frac{\mu}{T} \Big)  - E^{(F)\alpha} \Big],
\een
where the symbol $P^{\alpha \beta} = g^{\alpha \beta} + u^\alpha u^\beta$ stands for the projector orthogonal to the four-velocity $u^\alpha$.
As was pointed out in \cite{son09}, if we did not take into account the influence 
of the anomalies (i.e., $C_i=0$),  one can interpret the equation (\ref{ent}) as the relation describing the entropy production.
We ought to have that the right-hand side of it should be greater or equal to zero. This condition leads to the reqiurement imposed
on $ \sigma_F,~\sigma_{F \tB},~\sigma_{B F}$ and $\sigma_B$. It can be shown by the direct calculations, that the proviso $\p_\alpha s^\alpha \ge 0$,
will be satisfied if
\be
\sigma_F \ge 0, \qquad \sigma_F ~\sigma_B - \sigma_{F \tB} ~\sigma_{B F} \ge 0, \qquad \sigma_B \ge 0.
\ee
The above demands originate from the Sylvester criterion for the positivity of the quadratic form, to which one can rewrite the right-hand side
of the relation (\ref{ent}).  Moreover, we supposed positivity of
viscosity parameters $\eta$ and $\zeta$  \cite{landau1959} entering the formula for $\tau^{\alpha \beta}$.

Thus, the equation (\ref{ent}) can be interpreted as describing  the entropy production. Its right-hand side 
is greater or equal to zero, as required by the second law of thermodynamics. 
The presence of anomalies changes the situation drastically. The terms with $C_i\ne0$ can have either 
sign and, when negative, can even overcome the rest 
of the terms appearing in the equation (\ref{ent}) and thus spoil the positivity of entropy 
production. Therefore, the entropy flux $s^\alpha$, as well as, all the dissipative terms contributing to the transport current 
have  to be modified.

The most general modification of the entropy current, which comprises
standard dissipation terms, vorticity $\omega_\alpha = (1/2) \ep_{\alpha \beta \rho \delta} u^\beta  \p^\rho u^\delta$ 
 and the terms proportional to the magnetic components of the two $U(1)$-gauge fields are taken in the form 
\be \label{gibb}
s^\alpha = s u^\alpha - \frac{\mu}{T}~V_F^\alpha - \frac{\mu_d}{T}~V_B^\alpha+ D~\omega^\alpha + D_B~B^{(F)\alpha} + D_{\tB}~\tB^{(B)\alpha}.
\ee
The  dissipative contribution to the $U(1)$-gauge field currents are also modified by new transport coefficients $\xi$, $\xi_B$,
$\xi_d$, $\xi_\tB$, $\xi_{F \tB}$ and $\xi_{\tB F} $
\ben \nonumber \label{vf}
V_F^\alpha &=& - \sigma_F~\bigg[ T~P^{\alpha \beta}~\p_\beta \Big( \frac{\mu}{T} \Big) - E^{(F)\alpha} \bigg] 
- \sigma_{F \tB}~\bigg[ T~P^{\alpha \beta}~\p_\beta \Big( \frac{\mu_d}{T} \Big) - \tE^{(B)\alpha} \bigg]  + \xi~\omega^\alpha  \\ 
&+& \xi_B~B^{(F)\alpha} + \xi_{F \tB} \tB^{(B)\alpha},\\ \nonumber \label{vb}
V_B^\alpha &=& - \sigma_B~\bigg[ T~P^{\alpha \beta}~\p_\beta \Big( \frac{\mu_d}{T} \Big)  -\tE^{(B)\alpha} \bigg] 
- \sigma_{B F}~\bigg[ T~P^{\alpha \beta}~\p_\beta \Big( \frac{\mu}{T} \Big)  - E^{(F)\alpha} \bigg] + \xi_d~\omega^\alpha \\ 
&+& \xi_\tB~\tB^{(B)\alpha} + \xi_{\tB F} B^{(F)\alpha}.
\een
The unknown functions  $\xi,~\xi_d,~\xi_B,~\xi_\tB, \xi_{F \tB},~\xi_{\tB F}~D,~D_B,~D_\tB$   depend on $T$ and $\mu,~\mu_d$. Our aim is
to find the general formula for these new transport coefficients induced by the quantum anomalies.

Assuming that all the anomaly coefficients $C_i\ne 0$, one finds the conditions required 
for the positivity of $\p_\alpha s^\alpha$. During the direct
calculations of the divergence of the entropy current, one encounters
the derivatives of the vorticity $\p_\alpha \omega^\alpha$ to the vorticity $\omega^\alpha$ itself and similarly,
the  $\p_\alpha B^\alpha$ is related to $B^\alpha$. For our hydrodynamics (linear in the derivatives of velocity)
it is enough to find the required relations for the  ideal fluid. 
They  may be achieved by projecting the underlying equations 
of motion (\ref{e21})-(\ref{e23}) of the hydrodynamical model along two orthogonal directions.
Namely, along $u^\alpha$ and $P^\alpha_\beta = \delta^\alpha_\beta + u^\alpha u_\beta$. 
As a result we achieve the following relations for the ideal hydrodynamics ( i.e., with $\tau^{\alpha \beta} =0, ~V_F^\alpha=V_B^\alpha=0$)
\ben \label{eq219}
\p_\alpha \omega^\alpha &=& \frac{2 \omega_\alpha}{\ep + p}~\Big(
-\p^\alpha p + F^{\alpha \beta}~j_\beta (F) + B^{\alpha \beta} ~j_\beta (B) \Big),\\ \label{eq220}
\p_\alpha B^{(F)\alpha} &=& -2 \omega_\alpha E^{(F)\alpha} + \frac{B^{(F)}_{\alpha}}{\ep + p} \Big(
- \p^\alpha p + F^{\alpha \beta}~j_\beta (F) + B^{\alpha \beta} ~j_\beta (B) \Big)
,\\ \label{eq221}
\p_\alpha \tB^{(B)\alpha} &=& -2 \omega_\alpha \tE^{(B)\alpha} + \frac{\tB^{(B)}_{\alpha}}{\ep + p} \Big(
- \p^\alpha p + F^{\alpha \beta}~j_\beta (F) + B^{\alpha \beta} ~j_\beta (B) \Big).
\een

We  evaluate of $\p_\alpha s^\alpha$ with the help of (\ref{eq219})-(\ref{eq221}).
However,
the resulting expression comprises a number of terms containing  
$\omega^\alpha$,~$B^{(F)\alpha},$~$\tB^{(B)\alpha},$~$\omega^\alpha  E^{(F)}_{\alpha}$,
~$\omega^\alpha \tE^{(B)}_{\alpha},~E^{(F)}_{\alpha} B^{(F)\alpha},$\\
$\tE^{(B)}_{\alpha} \tB^{(B)\alpha}, ~\tE^{(B)}_{\alpha} B^{(F)\alpha},$ 
$E^{(F)}_{\alpha} \tB^{(B)\alpha} $.
The  condition $\p_\alpha s^\alpha \ge 0$ demands vanishing all factors multiplying the above terms. 
It eventuates in the following differential equations  
\ben \label{p19}
\p_\alpha D &-& 2 \frac{\p_\alpha p}{\ep + p} D - \xi~\p_\alpha \Big( \frac{\mu}{T} \Big)  - \xi_d~\p_\alpha \Big( \frac{\mu_d}{T} \Big)  = 0,\\ \label{p20}
\p_\alpha D_B &-& \frac{\p_\alpha p}{\ep + p} D_B - \xi_B~\p_\alpha \Big( \frac{\mu}{T} \Big)  
- \xi_{\tB F}~\p_\alpha \Big( \frac{\mu_d}{T} \Big)
= 0,\\ 
\label{p21}
\p_\alpha D_\tB &-& \frac{\p_\alpha p}{\ep + p} D_\tB - \xi_\tB~\p_\alpha \Big( \frac{\mu_d}{T} \Big)  - \xi_{F \tB}~\p_\alpha \Big( \frac{\mu}{T} \Big)= 0,
\een
and the additional conditions  
\ben
\label{p22}
\frac{2 D~\rho}{\ep + p} &-& 2  D_B + \frac{1}{T}~\xi = 0,\\ 
\label{p23}
\frac{2 D~\rho_d}{\ep + p} &-& 2 D_\tB + \frac{1}{T}~\xi_d = 0,\\ \label{p24}
\frac{\rho~D_B}{\ep + p} &+& \frac{\xi_B}{T} - \mu~\frac{C_1}{T} = 0,\\   
\label{p25}
\frac{\rho_d~D_\tB}{\ep + p} &+& \frac{\xi_\tB}{T} - \mu~\frac{C_2}{T} = 0,\\ 
\label{p26}
\frac{\rho~D_\tB}{\ep + p} &-& \mu_d~\frac{C_4}{T} + \frac{\xi_{F \tB}}{T} = 0,\\ \label{p27}
\frac{\rho_d~D_B}{\ep + p} &-& \mu_d~\frac{C_3}{T} + \frac{\xi_{\tB F}}{T} = 0.
\een
The differential equations (\ref{p19})-(\ref{p21}) suggest the dependence of the parameters $D_i = D,~ D_B,~ D_\tB$
on the pressure $p$ and the normalized chemical potentials $\tmu = \mu/T$ and $\tmu_d = \mu_d/T$. 
To exploit this fact we use thermodynamic relations
\ben
\p_\alpha D &=& \left(\frac{\p D}{\p p}\right)_{\tmu,\tmu_d} \p_\alpha p + \left(\frac{\p D}{\p \tmu}\right)_{p,\tmu_d} \p_\alpha \tmu + \left(\frac{\p D}{\p \tmu_d}\right)_{p,\tmu} \p_\alpha \tmu_d,\\
\p_\alpha D_B &=& \left(\frac{\p D_B}{\p p}\right)_{\tmu,\tmu_d} \p_\alpha p + \left(\frac{\p D_B}{\p \tmu}\right)_{p,\tmu_d} \p_\alpha \tmu + \left(\frac{\p D_B}{\p \tmu_d}\right)_{p,\tmu} \p_\alpha \tmu_d,\\
\p_\alpha D_\tB &=& \left(\frac{\p D_\tB}{\p p}\right)_{\tmu,\tmu_d} \p_\alpha p + \left(\frac{\p D_\tB}{\p \tmu}\right)_{p,\tmu_d} \p_\alpha \tmu + \left(\frac{\p D_\tB}{\p \tmu_d}\right)_{p,\tmu} \p_\alpha \tmu_d,
\een
and  require vanishing of the coefficients multiplying $\p_\alpha p$, $\p_\alpha \tmu$  and $\p_\alpha \tmu_d$, 
which can be considered as having arbitrary values at the initial time slice \cite{son09}. 
This leads to  three sets of the differential equations. The first defines the parameter $D(p,\tmu,\tmu_d)$
\ben \label{xi0}
\Big( \frac{\p D}{\p p}\Big)_{\tmu,\tmu_d} &-& \frac{2 D}{\ep + p} = 0, \\ \label{xi}
\Big( \frac{\p D}{\p \tmu} \Big)_{p,\tmu_d} &-& \xi = 0,\\ \label{xid}
\Big( \frac{\p D}{\p \tmu_d} \Big)_{p,\tmu}&-& \xi_d = 0,
\een
while the next two give the dependence of the partial derivatives of $D_B(p,\tmu,\tmu_d)$
\ben \label{xib0}
\Big( \frac{\p D_B}{\p p}\Big)_{\tmu,\tmu_d} &-& \frac{ D_B}{\ep + p} = 0,\\ \label{xib}
\Big( \frac{\p D_B}{\p \tmu} \Big)_{p,\tmu_d} &-& \xi_B = 0,\\ 
\Big( \frac{\p D_B}{\p \tmu_d}\Big)_{p,\tmu} &-& \xi_{\tB F} = 0,
\een
and $D_\tB(p,\tmu,\tmu_d)$
\ben \label{xitb0}
\Big( \frac{\p D_\tB}{\p p} \Big)_{\tmu_d,\tmu} &-& \frac{ D_\tB}{\ep + p} = 0,\\ \label{xitb1}
\Big( \frac{\p D_\tB}{\p \tmu} \Big)_{p,\tmu_d} &-& \xi_{F \tB} = 0,\\ \label{xitb}
\Big( \frac{\p D_\tB}{\p \tmu_d} \Big)_{p,\tmu} &-& \xi_\tB = 0. 
\een
Using the Gibbs-Duhem thermodynamic relations (\ref{gibbs}) we can arrive at the expression
\be
dp=\frac{\ep+p}{T}dT+\rho T d\tmu +\rho_d Td\tmu_d
\ee
which in turn can be easily cast into
\be
dT=\frac{T}{\ep+p}dp-\frac{\rho T^2}{\ep+p} d\tmu -\frac{\rho_d T^2}{\ep+p} d\tmu_d.
\ee
This provides the  relations as follows:
\be
\Big( \frac{\p T}{\p p} \Big)_{\tmu,\tmu_d} = \frac{T}{\ep + p}, \qquad \Big( \frac{\p T}{\p \tmu} \Big)_{p,\tmu_d} 
= - \frac{ \rho~T^2}{\ep + p}, \qquad \Big( \frac{\p T}{\p \tmu_d} \Big)_{p,\tmu} = - \frac{ \rho_d~T^2}{\ep + p}.
\label{threl}
\ee
 By virtue of (\ref{threl})
 the first equations from each of the sets of the relations (\ref{xi0}), (\ref{xib0}) and (\ref{xitb0}), can be immediately 
integrated. The results yields
\be
D=T^2d(\tmu,\tmu_d), \quad D_B=Td_B(\tmu,\tmu_d), \quad   D_\tB=Td_\tB(\tmu,\tmu_d), 
\label{int}
\ee
where
$d_i=d(\tmu,\tmu_d), ~d_B(\tmu,\tmu_d), ~d_\tB(\tmu,\tmu_d)$ are the new functions, which do not depend on temperature  $T$.
Thus it is more convenient to treat $D_i$ as functions of temperature $T$, and chemical potentials  $\tmu$ and $\tmu_d$.

To this end we assume the following dependence of the temperature $T=T(p,~\tmu,~\tmu_d)$ and use the relation
\be
\left(\frac{\p D_i(T,\tmu,\tmu_d)}{\p \tmu}\right)_{p,\tmu_d}=\left(\frac{\p D_i(T,\tmu,\tmu_d)}{\p \tmu}\right)_{T,\tmu_d}+
\left(\frac{\p D_i(T,\tmu,\tmu_d)}{\p T}\right)_{\tmu,\tmu_d}\left(\frac{\p T}{\p \tmu}\right)_{p,\tmu_d}.
\label{for}
\ee 
The formula similar to (\ref{for}) for the derivative with respect to $\tmu_d$ is supposed.
This leads to the system of differential equations provided by
\ben \label{p42}
T \Big( \frac{\p D}{\p T}\Big)_{\tmu,\tmu_d} &-& {2 D}= 0,\\ \label{p43}
\Big( \frac{\p D}{\p \tmu} \Big)_{T,\tmu_d}  &-& \frac{\rho~T^2}{\ep + p} ~\Big( \frac{\p D}{\p T} \Big)_{\tmu,\tmu_d} - \xi = 0,\\ \label{p44}
\Big( \frac{\p D}{\p \tmu_d} \Big)_{T,\tmu}&-&\frac{\rho_d~T^2}{\ep + p} ~\Big( \frac{\p D}{\p T} \Big)_{\tmu,\tmu_d} - \xi_d =0, 
\een
and for $D_B$ one gets
\ben \label{p45}
T \Big( \frac{\p D_B}{\p T}\Big)_{\tmu,\tmu_d} &-& D_B = 0,\\ \label{p46}
\Big( \frac{\p D_B}{\p \tmu} \Big)_{T,\tmu_d}  &-& \frac{\rho~T^2}{\ep + p} ~\Big( \frac{\p D_B}{\p T} \Big)_{\tmu.\tmu_d} 
- \xi_B = 0,\\  \label{p47}
\Big( \frac{\p D_B}{\p \tmu_d} \Big)_{T,\tmu}&-&\frac{\rho_d~T^2}{\ep + p} ~\Big( \frac{\p D}{\p T} \Big)_{\tmu,\tmu_d} - \xi_{\tB F} =0.
\een
Consequently, one obtains the similar equations for $D_\tB$
\ben \label{p48}
T \Big( \frac{\p D_\tB}{\p T}\Big)_{\tmu,\tmu_d} &-& D_\tB = 0,\\ \label{p49}
\Big( \frac{\p D_\tB}{\p \tmu} \Big)_{T,\tmu_d}  &-& \frac{\rho~T^2}{\ep + p} ~\Big( \frac{\p D_\tB}{\p T} \Big)_{\tmu,\tmu_d} - \xi_{F \tB} = 0,\\  \label{p50}
\Big( \frac{\p D_\tB}{\p \tmu_d} \Big)_{T,\tmu}&-&\frac{\rho_d~T^2}{\ep + p} ~\Big( \frac{\p D_\tB}{\p T} \Big)_{\tmu,\tmu_d} - \xi_\tB =0. 
\een
To proceed, we shall  replace  the derivatives of the type $\Big( \frac{\p D_i}{\p T}\Big)_{\tmu,\tmu_d}$, 
combining relations resulting from the inspections of
 (\ref{p42}),~(\ref{p45}),~(\ref{p48}),
 and inserting them into the adequate equations (\ref{p43})-( \ref{p44}),
(\ref{p46})-(\ref{p47}),~ (\ref{p49})-(\ref{p50}), respectively. Consequently, we obtain the three sets of partial differential equations
\ben
\Big( \frac{\p D}{\p \tmu} \Big)_{T,\tmu_d}&=&\frac{2\rho T}{\ep+p}D+\xi=2TD_B, \\
\Big( \frac{\p D}{\p \tmu_d} \Big)_{T,\tmu}&=&\frac{2\rho_d T}{\ep+p}D+\xi_d=2TD_\tB,
\een 
where the second equalities follow from the equations (\ref{p22}) and (\ref{p23})
\ben
\Big( \frac{\p D_B}{\p \tmu} \Big)_{T,\tmu_d}&=&\frac{\rho T}{\ep+p}D_B+\xi_B=C_1T\tmu, \\
\Big( \frac{\p D_B}{\p \tmu_d} \Big)_{T,\tmu}&=&\frac{\rho_d T}{\ep+p}D_B + \xi_{\tB F} =C_3T\tmu_d.
\een 
In the above derivations we use the relations (\ref{p24}) and (\ref{p27}). The last set of the equations can be easily achieved
by incorporating (\ref{p26}) and (\ref{p25}). Namely, one has
\ben
\Big( \frac{\p D_\tB}{\p \tmu} \Big)_{T,\tmu_d}&=&\frac{\rho T}{\ep+p}D_\tB + \xi_{F \tB} = C_4T\tmu_d, \\
\Big( \frac{\p D_\tB}{\p \tmu_d} \Big)_{T,\tmu}&=&\frac{\rho_d T}{\ep+p}D_\tB+\xi_\tB= C_2T\tmu.
\een 
The symmetry between the last two equations implies the equality $C_2=C_4$ and  in the following we shall
use the parameter $C_2$ only.
It is customary to write  
the solutions of the aforementioned sets of the partial differential equations as follows \cite{nei11} 
\ben 
D_B&=&\frac{1}{2}C_1T\tmu^2+\frac{1}{2}C_3T\tmu_d^2+\gamma_1 (T), \\ 
D_\tB&=& C_2 T\tmu~ \tmu_d+\gamma_2(T), \\ 
D&=&\frac{1}{3}C_1T^2\tmu^3 + C_2 T^2\tmu~ \tmu_d^2 + 2 \gamma_1(T) \tmu+ 2 \gamma_2 (T) \tmu_d+\gamma_3 (T),
\een
where $\gamma_i(T)$,~$i=1, 2, 3$ are temperature dependent functions. Using in the next step relations (\ref{p42}),~(\ref{p45}) and (\ref{p48}),
we achieve the following:
\ben \label{soldb}
D_B&=&\frac{1}{2}C_1T\tmu^2+\frac{1}{2}C_3 T\tmu_d^2+\tga_1 T, \\ \label{soltdb}
D_\tB&=& C_2 T\tmu~ \tmu_d+\tga_2~T, \\ \label{sold} 
D&=&\frac{1}{3}C_1T^2\tmu^3 + C_2 T^2\tmu~ \tmu_d^2 + 2 \tga_1T^2 \tmu + 2\tga_2 T^2 \tmu_d+\tga_3 T^2,
\een
where we have denoted by $\tga_i$,~$i=1, 2, 3$ integration constants.

Consequently, one can readily get the expressions for the novel
kinetic coefficients of the forms
\ben \label{eq267} \nonumber
\xi&=&C_1\mu^2\bigg(1-\frac{2}{3}\frac{\rho ~\mu}{\ep+p}\bigg) + \mu_d^2 C_2 \bigg( 1 - 2 \frac{\rho ~\mu}{\ep + p} \bigg)
- 4 \frac{\rho~T^2}{\ep + p} \bigg( \mu \tga_1 + \mu_d \tga_2 \bigg) \\
&-&  2 \frac{\rho ~T^3}{\ep + p} \tga_3 + 2 \tga_1 ~T^2, \\
\label{eq268} 
\xi_d&=& -\frac{2}{3}C_1\frac{\rho_d~ \mu^3}{\ep+p} + 2 C_2~ \mu~\mu_d \bigg(1-\frac{\rho_d~\mu_d}{\ep+p}\bigg) 
- 4 \frac{\rho_d~T^2}{\ep + p} \bigg( \mu \tga_1 + \mu_d \tga_2 \bigg) \\ \nonumber
&-&  2 \frac{\rho_d ~T^3}{\ep + p} \tga_3 + 2 \tga_2~ T^2,
\\
\label{eq269}
\xi_B&=& C_1 \mu  \bigg(1-\frac{1}{2}\frac{\rho~ \mu}{\ep+p}\bigg) - \frac{1}{2}C_3\frac{\rho ~\mu_d^2}{\ep+p} - \frac{\rho ~T^2}{\ep + p} \tga_1,\\
\label{eq270}
\xi_\tB&=& C_2 \mu \bigg(1-\frac{\rho_d ~\mu_d}{\ep+p}\bigg) - \frac{\rho_d~ T^2}{\ep + p} \tga_2.\\
\label{eq270a}
\xi_{F \tB} &=& C_2 \mu_d \bigg( 1 - \frac{\rho~ \mu}{\ep + p} \bigg) - \frac{\rho~T^2}{\ep + p} \tga_2,\\ \label{eq272}
\xi_{\tB F} &=& C_3 \mu_d \bigg( 1 - \frac{1}{2} \frac{\rho_d ~\mu_d}{\ep + p} \bigg) - \frac{1}{2} C_1 \frac{\rho_d ~\mu^2 }{\ep+p} 
- \frac{\rho_d~T^2}{\ep +p}~\tga_1.
\een
Equations (\ref{eq267})-(\ref{eq270b})  constitute the main results of the paper. They provide
the generalization and in the appropriate limit reduce to those obtained earlier \cite{son09}.


\section{Application to Dirac semi-metals with $\zz$ topological charge}
As was mentioned in the introduction most of the known Dirac semi-metals, in particular Na$_3$Bi or Cd$_2$As$_3$,
possess a chiral anomaly and two Dirac nodes, each carrying topological $\zz$  charge. In these materials two
Dirac nodes are protected by rotational symmetry of the crystal. The two anomalies show up in our results
as two different chemical potentials:  $\mu$ corresponds to the chiral anomaly and its change results
in the appearance of the chiral currents while $\mu_d$ decides about the position in energy of the two Dirac nodes.
In the presence of a magnetic field parallel to an electric field the corresponding currents are not conserved.
The current related to $\zz$ anomaly is a spin current, at least so, when the spin is approximately conserved \cite{burkov2016}. 

First, let us analyze the obtained results for the kinetic coefficients in the system in question, in the light of the analog of {\it chiral magnetic} and {\it chiral vortical} effects. Roughly 
speaking by {\it chiral magnetic} effect one understands the induction of electromagnetic current by means of an external magnetic field to chiral media with non-vanishing chemical potentials
\cite{kha06}.
On the other hand, the {\it chiral vortical} effect is bounded with an axial current in the direction of the local angular velocity \cite{kha11son}.


Let us commence 
with the analog of the {\it chiral magnetic effect} in the system under consideration. One has
two kinetic coefficients connected with $B^{(F)\mu}$, i.e., $\xi_{B}$ and $\xi_{\tB F}$. From the equation (\ref{eq269})
it can be seen that although the kinetic coefficient is connected with the {\it chiral magnetic} effect responsible for ordinary Maxwell field it contains the influence of the auxiliary $U(1)$-gauge field. 
The relation (\ref{eq272}) reveals also the dependence of both fields in the kinetic coefficient bounded with 
the {\it chiral vortical} effect. 
In the limit of square power of chemical potential and temperature tending to zero value, one arrives at 
the leading terms in {\it magnetic chiral} effect. Namely they yield
\be
\xi_B = C_1~\mu + \cO(\mu_m^{n \ge 2}, T), \qquad \xi_{\tB F} = C_3~\mu_d + \cO(\mu_m^{n \ge 2}, T),
\ee
and the leading term in {\it chiral vortical} effect is provided by
\be
\xi = C_1~\mu^2  + C_2~\mu_d^2 + \cO(\mu_m^{n \ge 3},T).
\ee
On the other hand, in the case of {\it chiral vortical} effect bounded with the auxiliary magnetic $U(1)$-gauge field $\tB^{(B)\mu}$,
there are two kinetic coefficients $\xi_{F \tB}$ and $\xi_{\tB}$
standing in front of the auxiliary magnetic field. Both equations (\ref{eq270a}) and (\ref{eq270}), reveal that in the case under consideration we have the influence of the ordinary Maxwell field, as well as the additional one. The same situation is  observed in $\xi_d$ coefficient, responsible for the {\it chiral vortical} effect. 
In the same limit as above, the leading terms describing {\it chiral magnetic} and {\it chiral vortical} effects connected with the auxiliary field are given, respectively by
\be
\xi_{F \tB} = C_2~\mu_d + \cO(\mu_m^{n \ge 2}, T), \qquad \xi_{\tB} = C_2~\mu + \cO(\mu_m^{n \ge 2}, T),
\ee
while the leading term in {\it chiral vortical} effect implies
\be
\xi_d = 2~C_2~\mu~\mu_d + \cO(\mu_m^{n \ge 3}, T).
\ee
Just in the system with chiral anomaly and $\zz$ anomalous topological charge we obtained four kinetic coefficients that exhibit {\it chiral magnetic} effect
$( \xi_B,~\xi_{\tB F},~\xi_{F \tB}, ~\xi_\tB)$, and two which determine {\it chiral vortical} effect $(\xi$,~$\xi_d)$.

Inspection of the equations defining currents $j_\alpha(F)$ and $j_\alpha (B)$, as well as, the definitions of $V_F^\alpha $ and $V_B^\alpha $ (\ref{vf})-(\ref{vb}),
reveal that the Maxwell field current is connected with the following kinetic coefficients $( \xi,~\xi_B,~\xi_{F \tB})$, while $j_\alpha(B)$ is bounded with 
$(\xi_d,~\xi_\tB,~\xi_{\tB F})$. In turn it leads to the conclusion that both $j_\alpha(F)$ and $j_\alpha (B)$ have terms depending on {\it chiral magnetic} effect, i.e.,
$j_\alpha(F) \sim (\xi_B,~\xi_{F \tB}),~j_\alpha (B) \sim (\xi_\tB,~\xi_{\tB F})$, and {\it chiral vortical} one, $j_\alpha(F) \sim \xi,~j_\alpha (B) \sim \xi_d$.
The result in question generalizes the relations on the currents obtained in \cite{burkov2016}, equations (13-14), because of the fact that the kinetic coefficients determined by the 
(\ref{eq267})-(\ref{eq272}) contain both the influence of the ordinary Maxwell field characteristics like $\rho,~\mu$, and the auxiliary $U(1)$-gauge field attributes, 
$\rho_d,~\mu_d$. The other bonus is the temperature dependence of them.

Let us give some remarks concerning the strength of both $U(1)$-gauge fields and their influence on the obtained kinetic coefficients.
In the case when $\rho$ and $\mu$ of the ordinary Maxwell field are far more greater than those for the auxiliary gauge field responsible for $\zz$ anomaly, we obtain
\ben 
\xi&=&C_1\mu^2\bigg(1-\frac{2}{3}\frac{\rho ~\mu}{\ep+p}\bigg) 
- 4 \frac{\rho~T^2}{\ep + p} ~ \mu \tga_1 -  2 \frac{\rho ~T^3}{\ep + p} \tga_3 + 2 \tga_1 ~T^2, \\
\xi_d&=&  2 \tga_2~ T^2,
\\
\xi_B&=& C_1 \mu  \bigg(1-\frac{1}{2}\frac{\rho~ \mu}{\ep+p}\bigg) - \frac{\rho ~T^2}{\ep + p} \tga_1,\\
\xi_\tB&=& C_2 \mu, \\
\xi_{F \tB} &=&  - \frac{\rho~T^2}{\ep + p} \tga_2,\\
\xi_{\tB F} &=& 0,
\een
On the other hand, in the opposite case when $\rho$ and $\mu$ tend to zero, one receives
\ben 
\xi&=& \mu_d^2~ C_2 - 4 \frac{\rho~T^2}{\ep + p}~\mu_d ~\tga_2  + 2 \tga_1 ~T^2, \\
\xi_d&=& - 4 \frac{\rho_d~T^2}{\ep + p}~ \mu_d \tga_2  -  2 \frac{\rho_d ~T^3}{\ep + p} \tga_3 + 2 \tga_2~ T^2,\\
\xi_B&=& 0,\\
\xi_\tB&=&  - \frac{\rho_d~ T^2}{\ep + p} \tga_2.\\
\xi_{F \tB} &=& C_2~ \mu_d,\\
\xi_{\tB F} &=& C_3 ~\mu_d \bigg( 1 - \frac{1}{2} \frac{\rho_d ~\mu_d}{\ep + p} \bigg) - \frac{1}{2} C_1 \frac{\rho_d ~\mu^2 }{\ep+p} 
- \frac{\rho_d~T^2}{\ep +p}~\tga_1.
\een
In the first limiting case, one can observe that up to the leading terms $ \cO(\mu_m^{n \ge 3}, T)$, the {\it chiral vortical} effect is connected with
$\xi \sim C_1 \mu^2$, while for the {\it chiral magnetic} effect, $\xi_B \sim C_1 \mu$ and $\xi_{\tB} \sim C_2 \mu$ kinetic coefficients, are responsible. 
On the other hand, when $\rho,~\mu \rightarrow 0$, the {\it chiral vortical} effect is bounded with $\xi \sim C_2 \mu_d^2$, and the {\it chiral magnetic}
effect is connected with $\xi_{F \tB} \sim C_2 \mu_d$ and $\xi_{\tB F} \sim C_3 \mu_d$.

Of course, one can see that both anomalies interplay will exert its influence on the magneto-transport properties, as was pointed out in \cite{burkov2016}.
The forms of the currents obtained in the aforementioned paper, i.e., their dependence on {\it chiral magnetic} and {\it vortical} effects played 
the crucial role in the experimentally
verifiable effects of magneto-transport.

In our case we obtained far more richer structure of the kinetic coefficients bounded with the adequate currents, so we conclude that the
magnetic conductivity will be modified by parameters of both considered fields and temperature.


Due to this interpretation of the $\zz$ bound current, one expects that spin - related magnetic field $\tB^{(B)}_{\alpha}$ vanishes \cite{burkov2016}.
This requirement causes that in the equation (\ref{vf}) there is no term with $\xi_{F \tB}$, and the relation (\ref{vb}) is lack of $\xi_{\tB}$.
Consequently we get the following kinetic coefficients:
\ben 
\xi&=&C_1\mu^2\bigg(1-\frac{2}{3}\frac{\rho ~\mu}{\ep+p}\bigg) + \mu_d^2 C_2 \bigg( 1 - 2 \frac{\rho ~\mu}{\ep + p} \bigg)
- 4 \frac{\rho~T^2}{\ep + p} \bigg( \mu \tga_1 + \mu_d \tga_2 \bigg) \\
&-&  2 \frac{\rho ~T^3}{\ep + p} \tga_3 + 2 \tga_1 ~T^2, \\
\xi_d&=& -\frac{2}{3}C_1\frac{\rho_d~ \mu^3}{\ep+p} + 2 C_2~ \mu~\mu_d \bigg(1-\frac{\rho_d~\mu_d}{\ep+p}\bigg) 
- 4 \frac{\rho_d~T^2}{\ep + p} \bigg( \mu \tga_1 + \mu_d \tga_2 \bigg) \\ \nonumber
&-&  2 \frac{\rho_d ~T^3}{\ep + p} \tga_3 + 2 \tga_2~ T^2,
\\
\xi_B&=& C_1 \mu  \bigg(1-\frac{1}{2}\frac{\rho~ \mu}{\ep+p}\bigg) - \frac{1}{2}C_3\frac{\rho ~\mu_d^2}{\ep+p} - \frac{\rho ~T^2}{\ep + p} \tga_1,\\
\xi_{\tB F} &=& C_3 \mu_d \bigg( 1 - \frac{1}{2} \frac{\rho_d ~\mu_d}{\ep + p} \bigg) - \frac{1}{2} C_1 \frac{\rho_d ~\mu^2 }{\ep+p} 
- \frac{\rho_d~T^2}{\ep +p}~\tga_1.
\label{eq270b}
\een
It is worth pointing out that even in the absence of $\tB^{(B)\alpha}$ field (but presence of $\tE^{(B) \alpha}$), one obtains two non-vanishing
kinetic coefficients $\xi$ and $\xi_d$, which exhibit the {\it chiral vortical} effect. They are functions of both chemical potentials and density currents, as well as, temperature.
In comparison to the case studied in \cite{son09}, one obtains $\xi_d$ kinetic coefficient, connected with spin conductivity (and possibly the spin Hall effect) which
is affected by both $U(1)$-gauge fields. Moreover the absence of $\tB^{(B)\alpha}$ induces the {\it chiral magnetic } effects, described by $\xi_B$ and $ \xi_{\tB F}$ coefficients.
In $\xi_B$, the $C_3$ parameter modifies the kinetic coefficient, which constitutes the novelty comparing to the previously studied case \cite{son09,nei11}. It contains
the influence of the auxiliary gauge field, its electric component.
The novelty is bounded also with $\xi_{\tB F}$, which mostly depends on $\rho_d,~\mu_d$, but also includes the influence of $\mu^2$.

 These findings harmonize with the recent kinetic calculations \cite{burkov2016},
where the authors have noted that the $\zz$ anomaly affects magneto-transport properties of Dirac semi-metals.
The observational manifestation of the $\zz$ anomaly found earlier is connected with the reduction of the diagonal resistivity due
to the spin Hall effect and the narrowing of the angular dependence of the magneto-resistance.
The detailed analysis of the magneto-conductivity and magneto-resistivity of the Weyl semi-metals based on the
presented hydrodynamic approach \cite{rog18} will be presented in the future publication.

\section{Summary and conclusions}
\label{sum-concl}
We have examined the generalized equations of relativistic hydrodynamics 
allowing the description of electrons in condensed matter systems with linear spectrum and the two different types of anomalies.
One of them is the well known chiral anomaly, while the  other one, authorizes the anomaly
observed in one class of Dirac semi-metal characterized by two Dirac nodes separated in momentum space
and lying on the axis of rotation.  With the  $\zz$ anomaly the corresponding charge density $\rho_d$ is connected. 
Its existence forces the non-trivial generalization of the relativistic hydrodynamics. 

We have found that the additional kinetic parameters, bounded with two different anomalous
charges and  required by the second law of thermodynamics 
and positiveness of the entropy production during the flow of electron fluid, enter the hydrodynamic 
equations in  the similar manner. They are a source of the additional kinetic coefficients called earlier
magnetic conductivities. In fact these are spin and spin Hall conductivities \cite{burkov2016}. 
Their appearance in the hydrodynamic equations can be traced back to 
 the necessity of adding dissipative terms proportional to the vorticity and magnetic components of the two $U(1)$-gauge fields.
Up to the first order in the velocity gradients,
they constitute the important component  in the proper description of the relativistic fluid. 

In the case under considerations  we obtain two kinetic coefficients exhibiting the {\it chiral vortical } effect, $\xi$ and $\xi_d$, and two
describing {\it chiral magnetic} effects. Interestingly the existence of $\zz$ anomaly induces kinetic coefficient $\xi_d$ connected with the $\zz$ related
conductivity. We argue that this conductivity is connected with spin conductivity and spin Hall effect in the kinetic approach to the problem in question.
The $\xi$ and $\xi_d$ coefficients are dependent on chemical potentials of both fields, their densities and temperature.

In the model under consideration, the absence of auxiliary magnetic field induces two kinetic coefficients bounded with the {\it chiral magnetic} effect 
($\xi,~\xi_{\tB F})$. In $\xi_d$ the existence of $C_3$ parameter modifies the previously obtained results. On the other hand, $\xi_{\tB F}$ depends mostly on 
the quantities characterizing the additional gauge field.
This
finding provides the generalization of the previous work on hydrodynamics with quantum triangle anomalies \cite{son09}. 

Let us discuss some possible experimental evidences of the theory in question. From the hydrodynamical calculations one obtains new formulae 
for the kinetic coefficients which constitute the vital ingredient of the currents connected with Maxwell and the auxiliary $U(1)$-gauge fields.
They reveal the dependence on temperature, {\it chiral magnetic} and {\it vortical} effects. On the other hand they are functions of characteristics of both fields.
In order to fully explore the validity of the hydrodynamic theory in real materials
being type II Dirac semimetals, one has to calculate the transport matrix
$\sigma_{(F, B,F\tB,BF)}$ for a system in weak electric and parallel to it magnetic
fields. The hydrodynamic theory provides a general information about the
positivity of the matrix, but to the given accuracy does not fix it. 

The possible experimental verifications of the presented theory might consist of the measurements (similar to those provided in \cite{goo17}) of the
dependence of magnetic conductivities on the parameters bounded with the gauge fields and 
constants $C_i$, related to the adequate type of anomaly. 
Such experiments might put some experimental restrictions on the aforementioned parameters.
Basing
on the kinetic theory presented in \cite{burkov2016}, one expects that additional anomaly
should lead to the narrowing of the magneto-conductivity curve. This will
be studied in the forthcoming paper, in which we shall calculate the
conductivity matrix $\sigma_{(F,B,F\tB,BF)}$ by hydrodynamic theory and
also using holography. The preliminary results show that in our approach
the presence of $\zz$ anomaly likewise leads to narrowing of the magneto-conductivity
line. The narrowing of the magneto-conductivity curve and its dependence on temperature were confirmed by the recent experimental data 
\cite{su2015, zha15, neupane2014,neupane2016}.


\acknowledgments
MR and KIW were partially supported by the grant DEC-2014/15/B/ST2/00089 of the National Science Center.
We would like to thank the Referee for valuable comments and remarks.



\end{document}